\documentclass[12pt,reqno]{article}
\usepackage{amsfonts}
\usepackage{amsmath}
\usepackage{amsbsy}
\usepackage{amssymb,latexsym}
\usepackage{setspace}
 \numberwithin{equation}{section}
\begin{document}
 \allowdisplaybreaks[1]
\title{Coset Algebras of the Maxwell-Einstein Supergravities}
\author{Nejat Tevfik Y$\i$lmaz\\
Department of Mathematics
and Computer Science,\\
\c{C}ankaya University,\\
\"{O}\u{g}retmenler Cad. No:14,\quad  06530,\\
 Balgat, Ankara, Turkey.\\
          \texttt{ntyilmaz@cankaya.edu.tr}}

\maketitle
\begin{abstract}
The general structure of the scalar cosets of the Maxwell-Einstein
supergravities is given. Following an introduction of the
non-linear coset formalism of the supergravity theories a
comparison of the coset algebras of the Maxwell-Einstein
supergravities in various dimensions is discussed.

\end{abstract}

\maketitle

\section{Introduction}
The Maxwell-Einstein supergravities are obtained by coupling an
arbitrary number of abelian vector multiplets to the supergravity
multiplet in various dimensions. They may also be obtained by the
Kaluza-Klein reduction on the Euclidean torus $T^{10-D}$ of the
ten dimensional simple $\mathcal{N}=1$ supergravity which is
coupled to $N$ abelian gauge multiplets \cite{heterotic}. The
scalar cosets of the Maxwell-Einstein supergarvity theories can be
formulated as non-linear sigma models, more specifically as the
symmetric space sigma models. The study of the scalar cosets of
these theories is essential in understanding the global symmetries
of the entire theory. The global symmetry of the scalar lagrangian
can be extended to the entire bosonic sector of the theory. The
scalar cosets $G/K$ of the Maxwell-Einstein theories are based on
the global internal symmetry groups $G$ which are in general
non-compact real forms of semi-simple Lie groups \cite{hel}. Under
certain conditions the global symmetry groups may be maximally
non-compact (split) real forms but in general they are elements of
a bigger class of Lie groups which contains the global symmetry
groups of the maximal supergravities \cite{julia1} namely the
split real forms as a special subset \cite{hel,julia2,ker1,ker2}.
The main difference between the scalar cosets based on the
non-compact and the maximally non-compact global symmetry groups
is the parametrization that one can choose for the coset
representatives. For the general non-compact real forms one can
make use of the solvable Lie algebra gauge \cite{fre} to
parameterize the scalar coset.

The Kaluza-Klein compactification of the bosonic sector of the ten
dimensional simple $\mathcal{N}=1$ supergravity which is coupled
to $N$ abelian gauge multiplets \cite{d=10} on the Euclidean torus
$T^{10-D}$ is given in \cite{heterotic}. When as a special case,
the number of the $U(1)$ gauge fields is chosen to be $16$, the
ten dimensional supergravity which is coupled to $16$ abelian
$U(1)$ gauge multiplets becomes the low energy effective limit of
the ten dimensional heterotic string theory. Thus the formulation
in this case corresponds to the dimensional reduction of the
low-energy effective bosonic lagrangian of the ten dimensional
heterotic string theory. When the number of coupling vector
multiplets is $N=16$, the $D=10$ Yang-Mills supergravity
\cite{d=10} has the $E_{8}\times E_{8}$ Yang-Mills gauge symmetry,
however the general Higgs vacuum structure causes a spontaneous
symmetry breakdown so that the full symmetry $E_{8}\times E_{8}$
is broken down to its maximal torus subgroup $U(1)^{16}$, whose
Lie algebra is the Cartan subalgebra of $E_{8}\times E_{8}$. Thus
the ten dimensional Yang-Mills supergravity reduces to its maximal
torus subtheory which is an abelian supergravity theory. The
bosonic sector of this abelian Yang-Mills supergravity corresponds
to the low energy effective Lagrangian of the bosonic sector of
the fully Higgsed ten dimensional heterotic string theory
\cite{d=10}. This abelian supergravity theory is the one where the
Maxwell-Einstein supergravities emerge from due to the
Kaluza-Klein reduction as we have discussed above.

The method of non-linear realizations
\cite{nej14,nej15,nej16,nej17,nej18} is used in
 \cite{nej19,nej110}
 to formulate the gravity as a non-linear realization in which the gravity
 and the gauge fields appear on equal footing. Later, the dualization of the bosonic fields
 has provided the non-linear realization formulation of the bosonic sectors of the maximal
supergravity theories \cite{julia1,julia2}. By introducing
auxiliary fields for a subset of the field content and
 by using the coset formulation, the global symmetries of the scalar sectors
 of the maximal supergravities are studied in detail in \cite{julia1}. These symmetries
 can also be realized on the bosonic fields. A general,
dimension-independent formalism is developed for the bosonic
sectors of the maximal supergravities in \cite{nej112}. The coset
 realizations
 of the non-gravitational bosonic sectors of the $D=11$ supergravity \cite{d=11}, the maximal supergravities which are obtained by
 the Kaluza-Klein reduction of the $D=11$ supergravity over the tori $T^{n}$, as well
as the IIB supergravity \cite{2B1,2B2,2B3}, are introduced by the
dualization of the scalars
 and the higher-order gauge fields in \cite{julia2}. In the same work, the
 twisted self-duality structure \cite{nej114,nej115} of the supergravities is generalized to
regain the first-order equations
 of the corresponding theories from the Cartan forms of the dualized coset.
 Therefore in \cite{julia2} it is shown that the
 non-linear coset formulation of the scalars can be improved to
 include the other non-gravitational bosonic fields, resulting in the first-order
 formulation of the relative theories. The mainline of \cite{julia2}
 is to introduce dual fields for the non-gravitational bosonic fields and
 to construct the Lie superalgebra which will generate the coset
 representatives that realize the original field equations both in first and second-order
 by means of the Cartan form of the coset map. The dualization method is another
 manifestation of the lagrange multiplier methods which are used for the scalar sectors of the
 maximal supergravities in \cite{julia1,pope}.

In this work we discuss the relative structures of the coset
algebras obtained as a result of the non-linear realization of the
bosonic sectors of the Maxwell-Einstein supergarvities in $D=7$
\cite{d=7}, $D=8$ \cite{d=8}, $D=9$ \cite{d=9}. We will first
mention about the general formulation of the scalar cosets of
these theories in section two. In section three after briefly
discussing the non-linear realization formalism we will give a
comparison of the coset algebras of the Maxwell-Einstein
supergravities which are constructed in \cite{nej5,nej4,nej6}
respectively.

\section{Scalar Cosets}
When the supergravity multiplet in $D$-dimensions is coupled to an
arbitrary number
 of $N$ abelian vector multiplets the scalars of the vector
 multiplets are governed by a symmetric space sigma model
 \cite{julia1,ker1,ker2,nej1,nej2}. The scalar fields $\varphi^{\alpha}$ for $\alpha=1,...,N(10-D)$ parameterize
the scalar coset manifold $SO(N,10-D)/SO(N)\times SO(10-D)$ where
$SO(N,10-D)$ is in general a non-compact real form of a
semi-simple Lie group and $SO(N)\times SO(10-D)$ is its maximal
compact subgroup. For this reason $SO(N,10-D)/SO(N)\times
SO(10-D)$ is a Riemannian globally symmetric space for all the
$SO(N,10-D)$-invariant Riemannian structures on
$SO(N,10-D)/SO(N)\times SO(10-D)$ \cite{hel}. Therefore the scalar
sector which consists of the vector multiplet scalars of the $D$
dimensional Maxwell-Einstein supergravity can be formulated as a
general symmetric space sigma model. To construct the symmetric
space sigma model lagrangian one may make use of the solvable Lie
algebra parametrization \cite{fre} for the parametrization of the
scalar coset manifold $SO(N,10-D)/SO(N)\times SO(10-D)$. The
solvable Lie algebra parametrization is a consequence of the
Iwasawa decomposition \cite{hel}
\begin{equation}\label{21}
\begin{aligned}
so(N,10-D)&=\mathbf{k}_{0}\oplus \mathbf{s}_{0}\\
\\&=\mathbf{k}_{0}\oplus \mathbf{h_{k}}\oplus
\mathbf{n_{k}},
\end{aligned}
\end{equation}
where $\mathbf{k}_{0}$ is the Lie algebra of $SO(N)\times
SO(10-D)$ and $\mathbf{s}_{0}$ is a solvable Lie subalgebra of
$so(N,10-D)$. In \eqref{21} $\mathbf{h}_{k}$ is a subalgebra of
the Cartan subalgebra $\mathbf{h}_{0}$ of $so(N,10-D)$ which
generates the maximal R-split torus in $SO(N,10-D)$
\cite{hel,ker2,nej2}. The nilpotent Lie subalgebra
$\mathbf{n_{k}}$ of $so(N,10-D)$ is generated by a subset
$\{E_{m}\}$ of the positive root generators of $so(N,10-D)$ where
$m\in\Delta_{nc}^{+}$. The roots in $\Delta_{nc}^{+}$ are the
non-compact roots with respect to the Cartan involution $\theta$
induced by the Cartan decomposition \cite{hel,nej2,nej3}
\begin{equation}\label{22}
so(N,10-D)=\mathbf{k}_{0}\oplus\mathbf{u}_{0},
\end{equation}
where $\mathbf{u}_{0}$ is a vector subspace of $so(N,10-D)$. By
using the scalar fields $\varphi^{\alpha}$ of the coupling vector
multiplets and the generators of the solvable Lie algebra
$\mathbf{s}_{0}$ we can parameterize the representatives of the
scalar coset manifold $SO(N,10-D)/SO(N)\times SO(10-D)$ as
\cite{hel}
\begin{equation}\label{23}
L=exp(\frac{1}{2}\phi ^{i}H_{i})exp(\chi ^{m}E_{m}),
\end{equation}
where $\{H_{i}\}$ for $i=1,...,$ dim($\mathbf{h}_{k})\equiv r$ are
the generators of $\mathbf{h}_{k}$ and $\{E_{m}\}$ for
$m\in\Delta_{nc}^{+}$ are the positive root generators which
generate $\mathbf{n_{k}}$. The scalars $\{\phi^{i}\}$ for
$i=1,...,r$ are called the dilatons and $\{\chi^{m}\}$ for
$m\in\Delta_{nc}^{+}$ are called the axions. The coset
representatives satisfy the defining relation of $SO(N,10-D)$
\begin{equation}\label{24}
L^{T}\eta L=\eta,
\end{equation}
where $\eta=$ diag$(-,-,...,-,+,+,...,+)$ in which there are
$10-D$ minus signs and $N$ plus signs. If we assume that we choose
the fundamental representation for the algebra $so(N,10-D)$ and if
we define the internal metric
\begin{equation}\label{25}
\mathcal{M}=L^{T}L,
\end{equation}
then the scalar lagrangian which governs the $N(10-D)$ scalar
fields of the vector multiplets can be constructed as
\begin{equation}\label{26}
 \mathcal{L}_{scalar}=\frac{1}{4}tr(\ast d\mathcal{M}^{-1}\wedge
 d\mathcal{M}).
\end{equation}

\section{Coset Formulation and the Coset Algebras}
In this section we will briefly mention about the general
formalism which formulates the bosonic sectors of the
Maxwell-Einstein supergravities as non-linear sigma models. As we
have discussed before the coset construction of the bosonic
sectors of the Maxwell-Einstein supergravities can be considered
to be an extension of the coset structure of the scalars of these
theories which we have given in the previous section. Such a
formulation treats the scalars and the other bosonic fields on
equal footing. One may apply the non-linear realization or the
dualization method of \cite{julia2} to construct the coset
formulation of the bosonic sectors of the Maxwell-Einstein
supergravity theories. Since the dualization method is another
manifestation of the langrange multiplier methods the bosonic
first-order formulation is also obtained as a consequence of the
coset construction. In such a coset formulation one first defines
a coset element which is generated by the bosonic fields coupled
to the generators of a Lie superalgebra. Then the Lie superalgebra
structure of the generators which parameterize this coset element
is derived so that the Cartan form induced by the coset map
realizes the field equations by satisfying the Cartan-Maurer
equation. The Cartan form of the dualized coset element will obey
a twisted self-duality equation \cite{julia2,julia3} which results
in the first-order bosonic field equations of the theory. To
construct the coset map the first task is to assign a generator
for each bosonic field. The original generators $\{T_{i}\}$ are
coupled to the fields $\{\tau^{i}\}$ in the coset parametrization.
One should also introduce a dual field for each original field.
The dual fields can be given as $\{\widetilde{\tau^{i}}\}$. These
dual fields are the langrange multipliers coupling to the Bianchi
identities of the field strengths of the original fields
\cite{pope}. Thus if the original field is a $p$-form the dual
field must be a $(D-p-2)-$form. We will also assign the dual
generators $\{\widetilde{T_{i}}\}$ to the dual fields so that they
will couple to the dual fields in the parametrization of the coset
element. The Lie superalgebra of the original and the dual
generators will have the $Z_{2}$ grading so that the generators
will be odd if the corresponding potential is an odd degree
differential form and otherwise even \cite{julia2}. Specifically
the doubled coset element will be parameterized by a differential
graded algebra. This algebra is generated by the differential
forms and the generators we have introduced above. The odd (even)
generators behave like odd (even) degree differential forms under
this graded differential algebra structure when they commute with
the differential forms. The odd generators obey the
anti-commutation relations while the even ones and the mixed ones
obey the commutation relations.

Apart from the graviton $e_{\mu}^{r}$ the bosonic field content of
the $D=7,8,9$ dimensional Maxwell-Einstein supergravities can be
given as
\begin{equation}\label{31}
(B_{\mu\nu},A_{\mu}^{I}, \sigma ,\varphi^{\alpha}),
\end{equation}
where $I=1,...,N+10-D$. The one-form fields $A_{\mu}^{I}$ include
the $N$ Maxwell fields of the vector multiplets and the $10-D$
vectors of the graviton multiplet. $B$ is a two-form field of the
graviton multiplet and $\sigma$ is the dilaton of the graviton
multiplet. The scalar fields $\varphi^{\alpha}$ belong to the
vector multiplets as mentioned before. The construction of the
coset formulation or the non-linear realization of the bosonic
sector of the Maxwell-Einstein supergravities requires the
definition of the coset element
\begin{equation}\label{32}
\begin{aligned}
\nu=&exp(\frac{1}{2}\phi^{j}H_{j})exp(\chi^{m}E_{m})exp(\sigma
K)exp(A^{I}V_{I})exp(\frac{1}{2}BY)\\
\\&\times
exp(\frac{1}{2}\widetilde{B}\widetilde{Y})exp(\widetilde{A}^{I}\widetilde{V}_{I})exp(\widetilde{\sigma}
\widetilde{K})exp(\widetilde{\chi}^{m}\widetilde{E}_{m})exp(\frac{1}{2}\widetilde{\phi}^{j}\widetilde{H}_{j}).
\end{aligned}
\end{equation}

Here we have defined the original generators
$\{K,V_{I},Y,H_{j},E_{m}\}$ and as we have mentioned above the
dual generators $\{\widetilde{K},\widetilde{V}_{I},\widetilde{Y},
\widetilde{H}_{j},\widetilde{E}_{m}\}$. The coset map $\nu$ is a
map from the $D$-dimensional spacetime into a group which is
presumably the rigid symmetry group of the dualized lagrangian.
However we will not focus on the group theoretical structure of
the non-linear realization of the Maxwell-Einstein supergravities
but rather on the Lie superalgebra which generates \eqref{32}.
Eventually this algebra also contains the information of the group
theoretical structure of the coset formulation
\cite{hel,julia1,ker1,ker2,nej1,nej2}. The local map $\nu$ induces
the Cartan form $\mathcal{G}$ on the $D$-dimensional spacetime
which can be given as
\begin{equation}\label{33}
\mathcal{G}=d\nu\nu^{-1}.
\end{equation}
From its construction the Cartan form \eqref{33} satisfies the
Cartan-Maurer equation
\begin{equation}\label{34}
d\mathcal{G}-\mathcal{G}\wedge\mathcal{G}=0.
\end{equation}
The standard dualization procedure
\cite{julia2,nej5,nej4,nej6,nej1,nej2} requires the construction
of the Lie superalgebra of the original and the dual generators
such that they will lead us to the second-order bosonic field
equations of motion when the Cartan form \eqref{33} is calculated
and inserted in \eqref{34}. Therefore as performed in
\cite{nej5,nej4,nej6} one can calculate the Cartan form \eqref{33}
in terms of the desired unknown structure constants of the Lie
superalgebra of the original and the dual generators then one can
insert this calculated Cartan form in the Cartan-Maurer equation
\eqref{34} and finally compare the result with the second-order
bosonic field equations to read the unknown structure constants.
This is the general method to determine the Lie superalgebra
structure which leads to the coset formulation of the
Maxwell-Einstein supergravities. Next we will present and discuss
the general structure of the coset algebras obtained as a result
of the above mentioned coset formulation of the $D=7$ \cite{d=7},
$D=8$ \cite{d=8}, $D=9$ \cite{d=9} Maxwell-Einstein
supergravities.

\subsection{The D=7 Case}
The bosonic lagrangian of the $\mathcal{N}=2$, $D=7$
Maxwell-Einstein supergravity can be given as \cite{d=7}
\begin{equation}\label{35}
\begin{aligned}
\mathcal{L}&=\frac{1}{2}R\ast1-\frac{5}{8} \ast d\sigma\wedge
d\sigma-\frac{1}{2}e^{2\sigma}\ast G
\wedge G\\
\\ &\quad -\frac{1}{8}tr( \ast d\mathcal{M}^{-1}\wedge
d\mathcal{M})-\frac{1}{2}e^{\sigma} F\wedge\mathcal{M} \ast F,
\end{aligned}
\end{equation}
where the coupling between the field strengths $F^{I}=dA^{I}$ for
$I=1,...,N+3$ and the scalars which parameterize the coset
$SO(N,3)/SO(N)\times SO(3)$ can be explicitly written as
\begin{equation}\label{36}
-\frac{1}{2}e^{\sigma} F\wedge\mathcal{M}\ast
F=-\frac{1}{2}e^{\sigma}\mathcal{M}_{ij} F^{i}\wedge \ast F^{j}.
\end{equation}
We have assumed the ($N+3$)-dimensional matrix representation of
$so(N,3)$. We have $\eta=$ diag$(-,-,-,+,+,...,+)$. The
Chern-Simons form $G$ is defined as \cite{d=7}
\begin{equation}\label{37}
 G=dB-\frac{1}{\sqrt{2}}\:\eta_{ij}\:A^{i}\wedge F^{j}.
\end{equation}
In \cite{nej5} the method of dualization which we have introduced
above is applied for the lagrangian \eqref{35} and the coset
algebra of the $\mathcal{N}=2$, $D=7$ Maxwell-Einstein
supergravity is found to be
\begin{eqnarray}
 [K,V_{i}]=\frac{1}{2}V_{i},\qquad[K,Y]=Y,\qquad[K,\widetilde{Y}]=-\widetilde{Y},\nonumber
\end{eqnarray}
\begin{eqnarray}
[\widetilde{V}_{k},K]=\frac{1}{2}\widetilde{V}_{k},\qquad\{V_{i},V_{j}\}=-\frac{1}{\sqrt{2}}\eta_{ij}Y,\qquad
[H_{l},V_{i}]=(H_{l})_{i}^{k}V_{k},\nonumber
\end{eqnarray}
\begin{eqnarray}
[E_{m},V_{i}]=(E_{m})_{i}^{j}V_{j},\qquad[V_{l},\widetilde{V}_{k}]=-\frac{2}{5}\delta_{lk}\widetilde{K}
+\frac{1}{2}\sum_{i=1}^{r}(H_{i})_{lk}\widetilde{H}_{i},\nonumber
\end{eqnarray}
\begin{eqnarray}
\{V_{k},\widetilde{Y}\}=2\sqrt{2}\,\eta_{k}^{l}\,\widetilde{V}_{l},\qquad[Y,\widetilde{Y}]=\frac{16}{5}\widetilde{K}
,\qquad[H_{i},\widetilde{V}_{k}]=-(H_{i}^{T})_{k}^{m}\widetilde{V}_{m},\nonumber
\end{eqnarray}
\begin{eqnarray}
[E_{\alpha},\widetilde{V}_{k}]=-(E_{\alpha}^{T})_{k}^{m}\widetilde{V}_{m},\qquad
[H_{j},E_{\alpha }]=\alpha _{j}E_{\alpha },\qquad [E_{\alpha
},E_{\beta }]=N_{\alpha ,\beta }E_{\alpha+\beta},\nonumber
\end{eqnarray}
\begin{eqnarray}
[H_{j},\widetilde{E}_{\alpha }]=-\alpha _{j}\widetilde{E}_{\alpha },\qquad [%
E_{\alpha },\widetilde{E}_{\alpha }]=\frac{1}{4}
\sum_{j=1}^{r}\alpha _{j}\widetilde{H}_{j},\nonumber
\end{eqnarray}
\begin{eqnarray}\label{38}
[E_{\alpha },\widetilde{E}_{\beta }]=N_{\alpha ,-\beta }\widetilde{E}%
_{\gamma },\qquad\alpha -\beta =-\gamma,\;\alpha \neq \beta,
\end{eqnarray}
where we have also included the commutation relations of the
generators of the solvable Lie subalgebra $\mathbf{s}_{0}$ of
$so(N,3)$ which form up a subalgebra in \eqref{38}. The matrices
($(H_{m})_{i}^{j}$, $(E_{\alpha})_{i}^{j}$) are the matrix
representatives of the corresponding generators
($H_{m},E_{\alpha}$). Also the matrices ($(H_{m}^{T})_{i}^{j}$,
$(E_{\alpha}^{T})_{i}^{j}$) are the matrix transpose of
($(H_{m})_{i}^{j}$, $(E_{\alpha})_{i}^{j}$). The commutators and
the anti-commutators which are not listed in \eqref{38} vanish.
\subsection{The D=8 Case}
The lagrangian of the bosonic sector of the $\mathcal{N}=1$, $D=8$
Maxwell-Einstein supergravity is given as \cite{d=8}
\begin{equation}\label{39}
\begin{aligned}
\mathcal{L}&=\frac{1}{4}R\ast1+\frac{3}{8} d\sigma\wedge \ast
d\sigma-\frac{1}{2}e^{2\sigma} G
\wedge \ast G\\
\\
&\quad+\frac{1}{16}tr( d\mathcal{M}^{-1}\wedge \ast
d\mathcal{M})-\frac{1}{2}e^{\sigma} F\wedge\mathcal{M} \ast F.
\end{aligned}
\end{equation}
The Chern-Simons three-form $G$ is
\begin{equation}\label{310}
 G=dB+\eta_{ij}F^{i}\wedge A^{j}.
\end{equation}
The scalars parameterize the coset manifold $SO(N,2)/SO(N)\times
SO(2)$ and the $SO(N,2)$ invariant tensor $\eta$ is $\eta=$
diag$(-,-,+,+,...,+)$ as generally defined before. We assume that
we choose an ($N+2$)-dimensional matrix representation of
$so(N,2)$. In \cite{nej4} the coset algebra which leads to the
non-linear sigma model or the coset formulation of the bosonic
sector of the $\mathcal{N}=1$, $D=8$ Maxwell-Einstein supergravity
is derived as
\begin{eqnarray}
[K,V_{i}]=\frac{1}{2}V_{i},\qquad[K,Y]=Y,\qquad[K,\widetilde{Y}]=-\widetilde{Y},\nonumber
\end{eqnarray}
\begin{eqnarray}
[\widetilde{V}_{k},K]=\frac{1}{2}\widetilde{V}_{k},\qquad\{V_{i},V_{j}\}=\eta_{ij}Y,\qquad
[H_{l},V_{i}]=(H_{l})_{i}^{k}V_{k},\nonumber
\end{eqnarray}
\begin{eqnarray}
[E_{m},V_{i}]=(E_{m})_{i}^{j}V_{j},\qquad\{V_{l},\widetilde{V}_{k}\}=\frac{2}{3}\delta_{lk}\widetilde{K}
+\sum_{i=1}^{r}(H_{i})_{lk}\widetilde{H}_{i},\nonumber
\end{eqnarray}
\begin{eqnarray}
[V_{k},\widetilde{Y}]=-4\,\eta_{k}^{l}\,\widetilde{V}_{l},\qquad[Y,\widetilde{Y}]=-\frac{16}{3}\widetilde{K},
\qquad[H_{i},\widetilde{V}_{k}]=-(H_{i}^{T})_{k}^{m}\widetilde{V}_{m},\nonumber
\end{eqnarray}
\begin{eqnarray}
[E_{\alpha},\widetilde{V}_{k}]=-(E_{\alpha}^{T})_{k}^{m}\widetilde{V}_{m},\qquad
[H_{j},E_{\alpha }]=\alpha _{j}E_{\alpha },\qquad [E_{\alpha
},E_{\beta }]=N_{\alpha ,\beta }E_{\alpha+\beta},\nonumber
\end{eqnarray}
\begin{eqnarray}
[H_{j},\widetilde{E}_{\alpha }]=-\alpha _{j}\widetilde{E}_{\alpha },\qquad [%
E_{\alpha },\widetilde{E}_{\alpha }]=\frac{1}{4}
\sum_{j=1}^{r}\alpha _{j}\widetilde{H}_{j},\nonumber
\end{eqnarray}
\begin{eqnarray}\label{311}
[E_{\alpha },\widetilde{E}_{\beta }]=N_{\alpha ,-\beta }\widetilde{E}%
_{\gamma },\qquad\alpha -\beta =-\gamma,\alpha \neq \beta.
\end{eqnarray}
As in the $D=7$ case the matrices ($(H_{m})_{i}^{j}$,
$(E_{\alpha})_{i}^{j}$) are the matrix representatives of the
corresponding generators ($H_{m},E_{\alpha}$) in the
($N+2$)-dimensional matrix representation of $so(N,2)$. Also the
matrices ($(H_{m}^{T})_{i}^{j}$, $(E_{\alpha}^{T})_{i}^{j}$) are
the matrix transpose of ($(H_{m})_{i}^{j}$,
$(E_{\alpha})_{i}^{j}$). The commutators and the anti-commutators
which are not listed in \eqref{311} vanish.
\subsection{The D=9 Case}
Finally we will present the coset algebra of the $\mathcal{N}=1$,
$D=9$ Maxwell-Einstein supergravity \cite{d=9} which is derived in
\cite{nej6}. The bosonic lagrangian of the $\mathcal{N}=1$, $D=9$
Maxwell-Einstein supergravity can be given as \cite{d=9}
\begin{equation}\label{312}
\begin{aligned}
\mathcal{L}&=-\frac{1}{4}R\ast1+\frac{7}{4} \ast d\sigma\wedge
d\sigma+\frac{1}{2}e^{-4\sigma}\ast G
\wedge G\\
\\
&\quad +\frac{1}{16}tr( \ast d\mathcal{M}^{-1}\wedge
d\mathcal{M})-\frac{1}{2}e^{-2\sigma} F\wedge\mathcal{M} \ast F.
\end{aligned}
\end{equation}
In this case the scalars of the coupling abelian vector multiplets
parameterize the scalar coset $SO(N,1)/SO(N)$. The Chern-Simons
three-form is taken as
\begin{equation}\label{313}
 G=dB+\eta_{IJ}A^{I}\wedge F^{J}.
\end{equation}
We assume an ($N+1$)-dimensional matrix representation of
$so(N,1)$. As before we have $\eta=$ diag$(-,+,+,...,+)$. The
coset algebra which parameterizes the coset \eqref{32} and which
generates the coset formulation of the bosonic sector is derived
in \cite{nej6} as
\begin{eqnarray}
[K,V_{I}]=-V_{I},\qquad[K,Y]=-2Y,\qquad[K,\widetilde{Y}]=2\widetilde{Y},\nonumber
\end{eqnarray}
\begin{eqnarray}
[\widetilde{V}_{I},K]=-\widetilde{V}_{I},\qquad\{V_{I},V_{J}\}=\eta_{IJ}Y,\qquad
[H_{l},V_{I}]=(H_{l})_{I}^{K}V_{K},\nonumber
\end{eqnarray}
\begin{eqnarray}
[E_{m},V_{I}]=(E_{m})_{I}^{J}V_{J},\qquad[V_{L},\widetilde{V}_{M}]=-\frac{2}{7}\delta_{LM}\widetilde{K}
-\sum_{i=1}^{r}(H_{i})_{LM}\widetilde{H}_{i},\nonumber
\end{eqnarray}
\begin{eqnarray}
\{V_{K},\widetilde{Y}\}=4\eta_{K}^{L}\,\widetilde{V}_{L},\qquad[Y,\widetilde{Y}]=-\frac{16}{7}\widetilde{K}
,\qquad[H_{i},\widetilde{V}_{K}]=-(H_{i}^{T})_{K}^{M}\widetilde{V}_{M},\nonumber
\end{eqnarray}
\begin{eqnarray}
[E_{\alpha},\widetilde{V}_{K}]=-(E_{\alpha}^{T})_{K}^{M}\widetilde{V}_{M},\qquad
[H_{j},E_{\alpha }]=\alpha _{j}E_{\alpha },\qquad [E_{\alpha
},E_{\beta }]=N_{\alpha ,\beta }E_{\alpha+\beta},\nonumber
\end{eqnarray}
\begin{eqnarray}
[H_{j},\widetilde{E}_{\alpha }]=-\alpha _{j}\widetilde{E}_{\alpha },\qquad [%
E_{\alpha },\widetilde{E}_{\alpha }]=\frac{1}{4}
\sum_{j=1}^{r}\alpha _{j}\widetilde{H}_{j},\nonumber
\end{eqnarray}
\begin{eqnarray}\label{314}
[E_{\alpha },\widetilde{E}_{\beta }]=N_{\alpha ,-\beta }\widetilde{E}%
_{\gamma },\qquad\alpha -\beta =-\gamma,\;\alpha \neq \beta.
\end{eqnarray}
As in the other cases the commutators and the anti-commutators
which are not listed above vanish.

We know that in general for the global symmetry algebra
$so(N,10-D)$ in the $D$-dimensional Maxwell-Einstein supergravity
the dimension of the solvable Lie algebra $ \mathbf{s}_{0}$ is
$N(10-D)$. Thus in a $D$-dimensional theory we have $N(10-D)$
dilatons and axions as the scalars of the vector multiplets which
parameterize the scalar coset $SO(N,10-D)/SO(N)\times SO(10-D)$.
From the coset algebras \eqref{38}, \eqref{311}, \eqref{314} we
also observe that these algebras contain the solvable Lie algebras
$ \mathbf{s}_{0}$ in them. One more observation is that the scalar
field generators and their duals form up a subalgebra in each
case. Thus one may think of the coset algebras as extensions of
the $2N(10-D)$ dimensional scalar-dual algebras in each case
whereas the scalar-dual algebras can be considered to be the
extensions of the solvable Lie algebras $ \mathbf{s}_{0}$ which
parameterize the scalar coset manifolds $SO(N,10-D)/SO(N)\times
SO(10-D)$. When we consider the bosonic field content of the
$D$-dimensional Maxwell-Einstein supergravities in general we have
a supergravity multiplet dilaton $\sigma$, a two form field $B$,
$N(10-D)$ vector multiplet scalars and we have $N+(10-D)$ one-form
fields. For the coset construction of the bosonic sector one has
to double the field content by introducing dual fields. Since we
also assign a generator for each field the dimension of the coset
algebra becomes
\begin{equation}\label{315}
\text{dim}(s_{dual})=22N-(2N+2)D+24.
\end{equation}
We deduce that the general scheme of the coset algebras in various
dimensions is the same however as a result of the coset
construction method we have discussed before, due to the oddness
or the evenness of the dimension $D$ of the spacetime the odd-even
structure of the generators differ. For all the coset algebras
given in \eqref{38}, \eqref{311} and \eqref{314} the generators
$\{\widetilde{K},\widetilde{H}_{i}\}$ commute with all the algebra
generators thus they generate the center $(s_{dual})_{c}$ of the
coset algebras $s_{dual}$ in each dimension. Therefore we have
\begin{equation}\label{316}
\text{dim}((s_{dual})_{c})=\text{dim}(\mathbf{h}_{k})+1.
\end{equation}
Finally we find that there are several abelian subalgebras of the
coset algebras $s_{dual}$ which are generated by the sets
\begin{eqnarray}
\{\widetilde{K},\widetilde{V}_{I},\widetilde{Y},
\widetilde{H}_{j},\widetilde{E}_{m}\},\quad
\{\widetilde{K},\widetilde{V}_{I}, Y,
\widetilde{H}_{j},\widetilde{E}_{m}\},\quad \{\widetilde{K}, Y,
\widetilde{H}_{j},H_{j}\},
\nonumber\\
\nonumber\\
\{\widetilde{K}, K, \widetilde{H}_{j},\widetilde{E}_{m}\},\quad
\{\widetilde{K}, K, \widetilde{H}_{j},H_{j}\},\quad
\{\widetilde{K}, H_{i},\widetilde{Y}, \widetilde{H}_{j}\},\quad\nonumber\\
\nonumber\\
 \{\widetilde{K}, H_{i}, Y,
 \widetilde{H}_{j}\}.\quad\quad\quad\quad\quad\quad\quad\quad\quad
 \label{316.5}
\end{eqnarray}
The dimension of the first two of these algebras which are maximal
in dimension is
\begin{equation}\label{317}
11N-(N+1)D+12,
\end{equation}
which is half of the dimension of the coset algebras $s_{dual}$.
The dimension of the abelian subalgebra generated by
$\{\widetilde{K}, K, \widetilde{H}_{j},\widetilde{E}_{m}\}$ is
\begin{equation}\label{318}
N(10-D)+2,
\end{equation}
and the dimension of the rest of the abelian subalgebras of
$s_{dual}$ which are given in \eqref{316.5} is
\begin{equation}\label{319}
2\:\text{dim}\mathbf{h_{k}}+2.
\end{equation}
As a final remark we can state that if the coset algebras
\eqref{38}, \eqref{311}, \eqref{314} form up a solvable Lie
algebra parametrization for a dualized coset structure
$G_{dual}/K_{dual}$ then the first two of the above mentioned
abelian algebras in \eqref{316.5} can be the candidates for the
subalgebra of the Cartan subalgebra of $g_{dual}$ which generates
the maximal R-split torus in $G_{dual}$.
\section{Conclusion}
In section two we have discussed the scalar coset structures of
the Maxwell-Einstein supergravities in general. After mentioning
the general formalism of the non-linear sigma model or the coset
formulation of the bosonic sectors of the Maxwell-Einstein
supergravity theories we have given a comparison of the coset
algebras derived in the coset constructions in various dimensions
in section three.

 The symmetries of the supergravity theories have
been studied in the recent years to gain insight in the symmetries
and the duality transformations of the string theories whose low
energy effective limits or the massless sectors are the
supergravities. The global symmetries of the supergravities help
us to understand the non-perturbative U-duality symmetries of the
string theories and the M theory \cite{nej125,nej126}. A
restriction of the global symmetry group $G$ of the supergravity
theory to the integers $Z$, namely $G(Z)$, is conjectured to be
the U-duality symmetry of the relative string theory which unifies
the T-duality and the S-duality \cite{nej125}. Therefore the coset
formulation of the supergravities have not only enabled us to
study the symmetries of the supergravities in detail but also
provided a better understanding of the dualities and the
symmetries of the string theories whose low energy effective
limits are the relative supergravities.

The Lie superalgebras we have presented in section three generate
the dualized coset elements. They may be considered as the
parametrization of a coset structure $G_{dual}/K_{dual}$ of the
bosonic sector likewise the coset structure $G/K$ of the scalars.
We know that the global (rigid) symmetries of the scalar sector
whose action on the scalar fields does not depend on the spacetime
coordinates are essential to have a deeper understanding of the
supergravity theories. One can also define the action of the
global symmetry group of the scalars on the other fields as well,
thus the global symmetry of the scalars can be extended to be the
 global symmetry of not only the bosonic sector but the entire theory.
The groups of the coset formulations namely $G_{dual}$ in various
dimensions can be studied as enlarged global symmetry groups of
the corresponding Maxwell-Einstein supergravity theories.
Therefore as we have mentioned before the improved global symmetry
analysis of the Maxwell-Einstein supergravities is an essential
tool to study the symmetry scheme of the relative heterotic string
theories since as discussed in the introduction the
Maxwell-Einstein supergravities are the low energy limits of the
heterotic string theories. In particular the
$T^{3}$-compactification of the $D=10$ type I supergravity that is
coupled to the Yang-Mills theory \cite{d=10,tani15} which forms up
the low energy effective limit of the $D=10$ heterotic string
gives the $D=7$ Maxwell-Einstein supergravity. Also an equivalent
dual bosonic lagrangian of the $D=7$ Maxwell-Einstein supergravity
in which the two-form potential $B$ is replaced by a dual
three-form field \cite{town1,town2} corresponds to the
$K_{3}$-compactification \cite{townref25} of the $D=11$
supergravity \cite{d=11} which is conjectured to be the low energy
limit of the M theory. Thus by constructing the coset algebra
which reveals information about the global symmetries of the $D=7$
Maxwell-Einstein supergravity we may have insight into the
symmetries of M theory. In \cite{town3} the string-membrane
dualities in $D=7$ which arise from the comparison of the
construction of the $D=7$ Maxwell-Einstein supergravity either as
a toroidally compactified heterotic string or a
$K_{3}$-compactified $D=11$ supermembrane are discussed. The coset
formulation of the bosonic sector of the $D=7$ Maxwell-Einstein
supergravity will also help to understand the string-membrane and
the string-string dualities in $D=7$.

The identification of the coset algebras we have given in section
three is not done. As a first guess one may consider them as
solvable algebras which are parts of Iwasawa decompositions
similar to the algebraic construction of the scalar cosets we have
mentioned in section two. For this reason one may focus on the
identification of the abelian and the nilpotent parts of
$s_{dual}$ by considering necessary generator redefinitions. The
relation of these algebras can be inspected with the coset
algebras of the $D=11$ and the maximal supergravities
\cite{julia2,west3,west4} as well. The attempt to identify the
coset structure $G_{dual}/K_{dual}$ and thus the enlarged global
symmetry group $G_{dual}$ must be in correspondence with the
identification of the coset algebras. However we should point out
that likewise proposed and applied in \cite{julia1} we make use of
a differential graded algebra of the module of the differential
forms and the algebra of the field generators for the
parametrization of the coset therefore the group theoretical
considerations of the coset formulation should posses structures
more involved than Lie groups. If one manages to construct the
group theoretical framework of the coset formulation one would
obtain a legitimate geometrical formulation of the related
supergravity theory at least for the bosonic sector. Another
essential inquiry may be held for the realization of the action of
the proposed global symmetry group $G_{dual}$ on the fields. In
\cite{julia2} it is argued that the symmetry groups generated by
the coset algebras become the symmetry groups of the Cartan forms
induced by the coset maps. Also they are conjectured to be the
symmetry groups of the first-order field equations obtained by a
twisted self-duality condition satisfied by the Cartan forms. One
may advance in this direction not only to identify the enlarged
global symmetry group $G_{dual}$ but also to explore the
transformation laws of its action on the field contents of the
related supergravities. If one manages to find out the nature of
$G_{dual}$ one may also work on the construction of a dualized
lagrangian which includes the original and the dual fields and
work out the action of $G_{dual}$ on the dualized lagrangian.

In \cite{westautobrane} the method of non-linear realizations is
used to derive the dynamics of the M theory branes. Furthermore
the M theory branes when they are in a background are also
described as non-linear realizations in \cite{west3}. Thus as a
final remark the comparison of the coset structures obtained in
these works with the coset algebras we have studied may reveal new
facts about the brane dynamics.

\end{document}